\documentclass[11pt]{article}
\usepackage{amsmath,amssymb,graphicx,hyperref,physics,listings,flafter}
\begin{document}
\begin{center}
\textbf{SIS/R model on Bi-Uniform hypergraphs}\\
\begin{tiny}
Rohan Nuckchady \\
\end{tiny}
\end{center}
\begin{center}
\textbf{Abstract}\\ 
\end{center}

This report is based on the work in (1). We first review definitions and notation developped there and provide derivations for the exact mathematical description of an SIS epidemic on a hypergraph. We then generalise the work in (1) to a new class of models that en-compass SIS and SIR models. The exact differential equations are derived for the expected values of the population of each state. Focusing on Bi-uniform hypergraphs, we make suitable approximations obtain numerical solutions to those equations. These are compared with stochastic simulations of the model for various systems.\\
\\
\textbf{Introduction}\\
\\
Models for epidemic propagation have usually been
 made on standard graphs. However these assume
pair-wise relationships between nodes, where as in reality
nodes interact in groups of more than 2. In order to account
for this, network based models often contain complete
subgraphs to model those larger groups, however these lead
to a loss of information in the system. Recently, work was
done in (1), (2) to develop epidemic models on hypergraphs.
Primarily SIS models were derived however these do not always provide a good model of epidemic propagation. In particular the type of hypergraph and relative sizes and number of hyper edges should heavily affect the reliability of the model. Often, epidemics propagate in a none-binary way-i.e there are more than two possible states per node. Such a model on graphs has been heavily studied in literature and is known as the SIR model. However, these assume that I-type nodes cannot transition back to S-type nodes. Hence these do not properly model epidemics where immunity is only gradually gained, or where the epidemic is deadly, infectious and where immunity is not guaranteed. This motivates the need to develop a model for these classes of epidemics. \\
\\
We begin by introducing basic ideas and definitions behind the model and reviewing the SIS model  in (1). We provide an exact mathematical description of the SIS model and derivations of the equations for the means of size of the population of each node type. Similar methods are then applied to derive equations for our new model (SISR). Our model will depend on two parameters and by adjusting their values both the SIS and SIR models can be recovered. Next, we describe the simulation methods used to produce stochastic simulations of the model on Bi-uniform hypergraphs. We also make a mean-field approximation valid for this class of hypergraphs. Finally, we compare the resultsof the stochastic simulations to numerical solutions to our
 system of equations.\\
\\
\textbf{Preliminaries}\\
\\
In this section we provide some basic definitions and develop the the SIS and SISR model on hypergraphs.\\
\\
\textbf{Definition 1.} $V$ is a set and its elements are called nodes. $E$ is another set whose elements consists of subsets of $V$ . The pair $(V, E)$ is called a hypergraph with vertices $v \in V$ and hyperedges $e \in E$.\\
\\
Thus in the case where $|e| = 2$ for all $e \in E$, we recover a 
graph. In the context of epidemic propagation, nodes on the hypergraph represent individuals and the hyperedges they are contained in represent the communities they are a member of. In the SIS model, each node can be in 2 possible states: i)Infected ii)Susceptible. There are two processes that are occuring in parallel on the hypergraph: i)Recovery ii) Infection. Recovery is the process by which an infected node becomes susceptible again and infection is the converse.\\
\\
\textbf{Definition 2.} $S$ denotes a susceptible node and $I$ an infected node. Infection and Recovery are modelled by poisson processes with parameters $r_{S}$ (node dependent) and $\gamma$ respectively such that the probability infection and recovery in a small amount of time $\delta t$ are given by $1-\exp \left(-r_{S} \delta t\right)$ and $1-\exp (-\gamma \delta t)$ respectively.\\
\\
Note with the above definition, $\gamma$ is taken to be independent of the nodes in the environment. It remains however to define how the environment (i.e. the hyperedges) affect the infection rate of a particular node. One could approximate a hypergraph by replacing each hyperedge with a complete subgraph. In doing so, the infection rate of a susceptible node would be the some $f(n)\tau$ , where $\tau$ is the infection rate per node, $n$ is the number of infected neighbours of the susceptible node and $f$ is an arbitrary function. In the case of a hypergraph though, each hyperedge containing the node provides its own infection pressure and thus we provide a different definition.\\
\\
\textbf{Definition 3.} Let $S$ be a susceptible node and $H_{S}=\{h \in$ $E \mid S \in h\}$. Let also $f: \mathbb{Z} \rightarrow \mathbb{R}$ and $\tau$ be the infection rate per node such that if a susceptible node has $n$ infected neighbours in a particular hyperedge, its infection rate is given by $f(n) \tau$. For $S$, the total infection rate $r_{S}$ is given by
$$
r_{S}=\sum_{h \in H_{S}} f\left(n_{h}\right)
$$
Where $n_{h}$ is the number of infectious nodes in $h \in E$.\\
\\
\textbf{Exact description of SIS model on hypergraphs}\\
\\

The state space for hypergraph with $|V|=N$ is $\{S, I\}^{N}$ containing $2^{N}$ elements. We partition this state space into $N+1$ classes.\\
\\
\textbf{Definition 4}. A class $S^{k}$ corresponds to all $\left(\begin{array}{l}N \\ k\end{array}\right)$ states containing $k$ infected nodes. $\left\{S^{k} \mid 0 \leq k \leq N\right\}$ is a partition of $\{S, I\}^{N} \cdot S_{i}^{k} \in S^{k}$ are indered elements of $S^{k}$ with $1 \leq i \leq$ $\left(\begin{array}{l}N \\ k\end{array}\right)=c_{k} . S_{i}^{k}(j) \in\{S, I\}$ denotes the state of the $j^{\text {th }}$ node.\\
\\
In this language we describe the processes of infection and recovery transitions:\\
i) Infection: An $S_{j}^{k} \rightarrow s_{i}^{k+1}$ transition, such that for some $p \in\{1, \ldots, N\}$, we have $S_{j}^{k}(p)=S, S_{i}^{k}(p)=I$ and otherwise $S_{j}^{k}(m)=S_{i}^{k-1}(m)$ for all $m \neq p, m \in\{1, \ldots, N\} .$ Also, there is a $q \neq p$ such that $S_{j}^{k}(q)=I$ and $p, q$ are both contained in the same hyperedge.
ii)Recovery: An $S_{j}^{k} \rightarrow S_{i}^{k-1}$ transition, such that there is some $p \in\{1, \ldots, N\}$ with $S_{j}^{k}(p)=I, S_{i}^{k-1}(p)=S$ and for all $m \neq p, m \in\{1, \ldots, N\}$ we have $S_{j}^{k}(m)=s_{i}^{k-1}(m)$\\
\\
These give rise to a linear system of differential equations for the vectors of probabilities $X^{k}(t)=\left(X_{1}^{k}, \ldots, X_{r_{k}}^{k}\right)$. Each $X_{i}^{k}$ denotes the probability that the system is in the state $S_{i}^{k}$ at time $t$. In particular this differential equation can be written as:
$$
\frac{d X^{k}}{d t}=A^{k} X^{k-1}+B^{k} X^{k}+C^{k} X^{k+1}
$$
$A^{k}$ describes recovery, $B^{k}$ describes the behaviour if nothing happens and $C^{k}$ describes infection. These can be written in the form of a Kolomogrov equation as $\frac{d X}{d t}=P X$, where $P$ is in block tri-diagonal form:
$$
P=\left(\begin{array}{ccccccc}
B^{0} & C^{0} & 0 & 0 & 0 & \cdots & 0 \\
A^{1} & B^{1} & C^{1} & 0 & 0 & \cdots & \vdots \\
0 & A^{2} & B^{2} & C^{2} & 0 & \cdots & \vdots \\
0 & 0 & A^{3} & B^{3} & C^{3} & \vdots & \vdots \\
\vdots & \vdots & \cdots & \cdots & \cdots & \vdots & \vdots \\
0 & 0 \ldots & \cdots & \cdots & \cdots & A^{n} & B^{N}
\end{array}\right)
$$
The matrix element $A_{i j}^{k}$ is the rate of transition from $S_{j}^{k-1}$ to $S_{i}^{k}$ and is therefore given by:
$$ A_{i j}^{k}= \begin{cases}\tau \sum_{h_{i \in h} h} f\left(N_{h}\left(S_{j}^{k-1}\right)\right) & \text { if an } S_{j}^{k-1} \rightarrow S_{1}^{k} \\ & \text { transition exists and } l \text { is the } \\ 0 & \text { newly infected node } \\ & \text { otherwise }\end{cases}$$
With $N_{h}$ denoting the number of infected nodes in $h$.\\
\\
Similarly $C_{i j}^{k}$ is the transition rate from $S_{j}^{k}$ to $S_{j-1}^{k-1}$ and
$$
C_{i j}^{k}=\left\{\begin{array}{l}
\gamma \quad \text { if an } S_{j}^{k-1} \rightarrow S_{i}^{k} \\
\quad \text { transition exints and } l \text { is the } \\
\quad \text { previously infected node } \\
0 \quad \text { otherwise }
\end{array}\right.
$$
$B^{k}$ describes transitions between $S_{j}^{k}$ and $S_{i}^{k}$, which is only possible in our model if $i=j$. Since the sum of elements in each column of $P$ is zero (3), then it follows that the diagonal elements of $B^{k}$ are given by
$$
B_{u}^{k}=-\sum_{\lambda I H} A_{m u}^{k+1}-\sum_{A H} C_{m u}^{*-1}
$$
We denote the first term in this equation by $N_{S t}$, and note that it represents the sum of $\sum_{\mathrm{A}=\mathrm{n}} f\left(N_{\mathrm{A}}\left(S_{j}^{k-1}\right)\right.$ over all susceptible nodes in $S_{j}^{k-1}$. The second term can be evaluated as $\gamma(k+1)$ since the number of non zero terms equals to the number of infected nodes in $S_{\mathrm{m}}^{\text {k }+1}$.
$$
B_{i j}^{k}= \begin{cases}N_{S u}-\gamma(k+1) & \text { if } i=j \\ 0 & \text { otherwise }\end{cases}
$$
While these provide the exact evolution of the system, it is very difficult to solve exactly and numerically for large hypergraphs. We therefore focus on developing approximations for the expected values of the number of infected and susceptible nodes.\\
\\
\textbf{Theorem 5} Let $[X]=\mathbb{E}(X)$ where $X=l$ ar $S$ and $[S I]=\sum_{k=0}^{N} \sum_{j=1}^{c} N_{S t} X_{j}^{k}(t)$ with $\alpha=\left(\begin{array}{l}N \\ k\end{array}\right)$. Then the following equations hold for arbitrary hypergraphs:
$$
\begin{aligned}
&\frac{d[S]}{d t}=\gamma[I]-\tau[S I] \\
&\frac{d[I]}{d t}=-\gamma[I]+\tau[S I]
\end{aligned}
$$
Proof. Introduce $c_{k}=(1,1, \ldots, 1)$ as vector with $\alpha$ ones. We can then write $[l]=\sum_{k=0}^{N} k e_{k} X^{k}$ and $|S|=\sum_{k=0}^{N}(N-k) e_{k} X^{k}$. We can also write 4 as $B_{4}^{k}=-\left(e_{k+1} A^{k+1}\right)_{i}-\left(e_{k-1} C^{k-1}\right)_{i}=$ $\left(c_{i} B^{k}\right)_{i}$. So it follows that
$$
a_{k+1} A^{k+1}+c_{k-1} C^{k-1}+c_{k} B^{k}=0
$$
Next we compute
$$
\begin{aligned}
\frac{d[S]}{d t} &=\sum_{k=0}^{N}(N-k) e_{k} \frac{d X^{k}}{d t} \\
&=\sum_{k=0}^{N}(N-k)\left(e_{k} A^{k} X^{k-1}+e_{k} B^{k} X^{k}+e_{k} C^{k} X^{k+1}\right) \\
&=\sum_{k=0}^{N}\left((N-k-1) e_{k+1} A^{k+1}+(N-k) e_{k} B^{k}\right.\\
&\left.+(N-k+1) C^{k-1}\right) X^{k} \\
&=-\sum_{k=0}^{N}\left(e_{k+1} A^{k+1}-e_{k-1} C^{k-1}\right) X^{k}
\end{aligned}
$$
In the first line of the above simplification we used 1, in the second line we re-labled the dummy variables and in the third line we used 8 twice to simplify. To continue we note as in 5 that $\left(e_{k-1} C^{k-1}\right)_{4}=\gamma_{k}=\left(\gamma k_{u}\right)_{2}$. Also note that $\left(e_{k+1} A^{k+1}\right)_{t}=\sum_{m=0}^{k+1} A_{m i}^{k+1}=F N_{S i}$ so that $\sum_{k=0}^{N} e_{k+1} A^{k+1} X^{k}=$ $r \sum_{k=0}^{N} N_{s Z} X^{*}=r|S T|$. Combining them we obtain
$$
\frac{d[S]}{d t}=\gamma[l]-r[S I]
$$
The proof of the second equation is similar and is done in (1). Alternatively one could note that $\mathbb{E}(I+S)=N$ and thus that $\frac{d[I]}{d t}=-\frac{d[S]}{d t}$.\\
\\
It is important to note at this stage that these equations do not provide a closed set of differential equations as $[S I]$ is not known. Various approximations can be made to estimate [SI] depending on the structure of the hypergraph and the form of $f$.\\
\\
\textbf{Exact description of SISR model on hypergraphs.}\\
\\
We introduce a new possible state for each node denoted as D. Our state space now has dimension $3^{N}$. We keep the same notation as in the $S I S$ model to denote our state space and introduce a new transition 'deactivation':\\
\\
\textbf{Definition 6.} The state space in the SISR model on a hypergraph with $|V|=N$ is $\{S, I, D\}^{N}$. We keep the notation of $0.4$ and introduce a class $R^{k}$ which correspond to all $2^{N-k}\left(\begin{array}{c}N \\ k\end{array}\right)=b_{k}$ states containing $k$ susceptible nodes. $R_{i}^{k} \in R^{k}$ are indexed elements of $R^{k}$ with $1 \leq i \leq c_{k} . R_{i}^{k}(j) \in\{S, I, D\}$ denotes the state of the $j^{\text {th }}$ node.\\
\\
The transitions in the model are identical to the SIS model but we add another a deactivation transition: An $S_{j}^{k} \rightarrow S_{i}^{k-1}$ transition, such that there exists a $p \in\{1, \ldots, N\}$ satisfying $S_{j}^{k}(p)=I$ and $S_{i}^{k-1}(p)=D$ and $S_{j}^{k}(m)=S_{m}^{k-1}$ for all $m \neq p, m \in\{1, \ldots, N\}$.
\\
\\
\textbf{Definition 7}. A deactivation is a poisson process between $S$ and $D$ with parameter $\beta$ such that the probability of deactivation in a small time $\delta t$ is $1-\exp (-\beta \delta t)$\\
\\

Next we introduce $Y^{k}=\left(Y_{1}^{k}, \ldots, Y_{c_{k}}^{k}\right)$ to denote the vector of probabilities with $Y_{i}^{k}$ being the probability that the system is in state $R_{i}^{k}$. We now have a pair of differential equations:
$$
\begin{aligned}
&\frac{d X^{k}}{d t}=A_{1}^{k} X^{k-1}+B_{1}^{k} X^{k}+C_{1}^{k} X^{k+1} \\
&\frac{d Y^{k}}{d t}=M^{k} Y^{k-1}+N^{k} Y^{k}+L^{k} Y^{k+1}
\end{aligned}
$$
$A_{1}$ again describes the rate of infection, so $A_{1}^{k}=A^{k}$ with some abuse of notation. $C_{1}$ describes recovery and deactivation processes, both of which correspond to $S^{k+1} \rightarrow S^{k}$ processes. We have
$$
C_{i j}^{k}= \begin{cases}\gamma & \text { if an } S_{j}^{k-1} \rightarrow S_{i}^{k} \text { infection } \\ & \text { transition exists. } \\ \beta & \text { if an } S_{j}^{k-1} \rightarrow S_{i}^{k} \text { deactivation } \\ & \text { transition exists. } \\ 0 & \text { otherwise }\end{cases}
$$
The matrix $B_{1}^{k}$ is determined identically as in the case of the SIS model.
$M_{i j}^{k}$ is the recovery rate from $R_{j}^{k-1}$ to $R_{i}^{k}$, so we have
$$
M_{i j}^{k}= \begin{cases}\gamma & \text { If a recovery transition } R_{j}^{k-1} t o R_{i}^{k} \\ & \text { exists } \\ 0 & \text { otherwise }\end{cases}
$$
$L_{i j}^{k}$ describes either infection of a single node. Hence we have;
$$
L_{i j}^{k}= \begin{cases}\tau \sum_{h: l \in h} f\left(N_{h}\left(R_{j}^{k+1}\right)\right) & \text { If an infection transition } \\ & \text { is possible and where } l \text { is the } \\ & \text { newly infected node } \\ 0 & \text { otherwise }\end{cases}
$$
The matrix $N^{k}$ is determined as in the case of the SIS model.
This system has higher dimension than the SIS model and we therefore focus on expected valued. We denote $[S I]_{t}=\mathbb{E}(S I)$\\
\\

\textbf{Theorem 6.} With all the definitions of $0.4$ and $[D]=$ $\mathbb{E}(D)$ the following hold for an arbitrary hypergraph:
$$
\frac{d[I]}{d t}=\tau[S I]-(\gamma+\beta)[I]
$$
$$
\begin{gathered}
\frac{d[S]}{d t}=\gamma[I]-\tau[S I] \\
\frac{d[D]}{d t}=\beta[I]
\end{gathered}
$$
Note that the evolution equation for $[S]$ is as expected, the same as in the SIS model.\\
\\
Proof. The proof is similar to that of the proof of $Theorem \, 5 $ or as done in (1). We will only show how the additional $\beta$ factor is obtained. This difference arises in the evaluation of $\left(e_{k-1} C^{k-1}\right)_{j}=\sum_{i} C_{i j}$. for fixed $j, C_{i j}^{k}$ is none zero $k$ times as $S_{j}^{k}$ has $k$ infected nodes. Each one of these can undergo the same transitions and each correspond to a unique term in $S^{k-1}$. So $\left(e_{k-1} C^{k-1}\right)_{j}=\left(k(\gamma+\beta) e_{k}\right)_{j}$.\\
\\
The proof 16 is similar, but we reproduce it here due to the different notation. We introduce $e_{k}=(1, \ldots, 1)$ with $b_{k}$ ones and similarly to the SIS model we have: $N_{i i}^{k}=\left(e_{k} N^{k}\right)_{i}=$ $-\left(e_{k+1} M^{k+1}\right)_{i}-\left(e_{k-1} L^{k-1}\right)_{i}$. Computing the derivative:
$$
\begin{aligned}
\frac{d[S]}{d t} &=\sum_{k=0}^{N} k e_{k} \frac{d Y^{k}}{d t} \\
&=\sum_{k=0}^{N}\left(e_{k+1} M^{k+1}-e_{k-1} L^{k-1}\right) Y^{k}
\end{aligned}
$$
The manipulations required are identical to the proof of $0.4 .1$. Now we look at each term separately. Let $N_{I}(G)$ denote the number of infected nodes in state $G$.
$$
\begin{aligned}
\sum_{k=0}^{N} e_{k+1} M^{k+1} Y^{k} &=\sum_{k=0}^{N} \sum_{a l l j} \sum_{a l l i} M_{i j}^{k+1} Y_{j}^{k} \\
&=\sum_{k=0}^{N} \sum_{a l l j} \gamma N_{I}\left(R_{j}^{k}\right) Y_{j}^{k} \\
&=\gamma[I]
\end{aligned}
$$
With $N_{I}$ denoting the number of infected nodes in the state. The last equality follows from the definition of expectation values in two state system since the classes $Y^{k}$ form a partition of the state space.
$$
\begin{aligned}
\sum_{k=0}^{N} e_{k-1} L^{k-1} Y^{k} &=\sum_{k=0}^{N} \sum_{a l l j} \sum_{a l l i} L_{i j}^{k-1} Y_{j}^{k} \\
&=\sum_{k=0}^{N} \sum_{a l l j} \sum_{a l l i} \tau \sum_{h: l \in h} f\left(N_{h}\left(R_{j}^{k}\right)\right) Y_{j}^{k} \\
&=\sum_{k=0}^{N} \sum_{a l l j} \tau N_{S I} Y_{j}^{k} \\
&=\tau[S I]
\end{aligned}
$$
The last two equality's follow because there is a bijection between states $R_{j}^{k}$ and $S_{j}^{k}$. This gives
$$
\frac{d[S]}{d t}=\gamma[I]-\tau[S I]
$$
Finally since $\mathbb{E}(S+I+D)=N$, then using linearity and differentiating, $\frac{d[D]}{d t}=-\frac{d[I]}{d t}-\frac{d[S]}{d t}=\beta[I]$\\
\\
\textbf{Simulation Methods}\\
\\
So far our discussion has mostly general, however for the aim of simulating epidemics, we choose to focus on Bi-Uniform Hypergraphs: Each vertex is contained in exactly two edges of two distinct types which we call their respective edge sets $H$ and $W$. Edges of type $H$ (respectively $W$ ) will be disjoint and form a partition of $V$ and $(V, H \cup W)$ is a hypergraph. All elements $h \in H$ have the same size $|h|=d_{h}$ and elements $w \in W$ also have the same size $|w|=d_{w} .$ Random generation of such a hypergraph is done as follows:
\\
\\ i) First generate a list $V$ of vertices. To select $k$ random vertices from $V$, we randomly order $V$ and pick the first $k$ elements. i) Generating $H$ : Given $d_{h}$, we thus randomly pick $d_{h}$ elements from $V$, then $d_{h}$ elements from $V / h_{1}$ and iterate so that at the $i^{t h}$ step we randomly select $d_{h}$ elements from $V \backslash\left(\cup_{j<i} h_{j}\right)$.\\
\\
ii) Generating $W$ : The same procedure as in generating $H$ is used.\\
Note that this requires $|V|$ to be divisible by $d_{h}$ and $d_{w}$ and we arbitrarily choose $d_{w} \geq d_{h}$
$f$ is chosen to be non linear, as otherwise it would be possible to get a very good approximation by using a complete graph model of the hypergraph:
$$
f=d_{w} \arctan \left(\frac{\pi x}{2\left(d_{w}-1\right)}\right)
$$
We make various approximations throughout to estimate $[S I]$. If the number total number of infected people is $I$, the number of infected nodes in $h \in H$ is approximated as $\frac{d_{h}-1}{N} I$. Similarly the number of infected nodes in $w \in W$ is approximated as $\frac{d_{w}-1}{N} I$. In the case of the $S I R$ model this leads to:
$$
\begin{aligned}
[S I](t) &=\sum_{k=0}^{N} \sum_{j=1}^{b_{k}} \sum_{l: R_{j}^{k}(l)=S} \sum_{h: l \in h} f\left(N_{h}\left(R_{j}^{k}\right)\right) Y_{j}^{k} \\
&=\sum_{k=0}^{N} \sum_{j=1}^{b_{k}} \sum_{l: R_{j}^{k}(l)=S} f\left(\frac{d_{w}-1}{N} I\right)+f\left(\frac{d_{h}-1}{N} I\right) Y_{j}^{k} \\
&=\sum_{k=0}^{N} \sum_{j=1}^{b_{k}} k f\left(\frac{d_{w}-1}{N} I\right)+f\left(\frac{d_{h}-1}{N} I\right) Y_{j}^{k} \\
&=\mathbb{E}\left(S(t)\left(f\left(\frac{d_{w}-1}{N} I(t)\right)+f\left(\frac{d_{h}-1}{N} I(t)\right)\right)\right) \\
&=[S]\left(f\left(\frac{d_{w}-1}{N}[I]\right)+f\left(\frac{d_{h}-1}{N}[I]\right)\right)
\end{aligned}
$$
In the last line we further approximated $S(t)$ to be independent of $f\left(\frac{d_{w}-1}{N} I\right)+f\left(\frac{d_{h}-1}{N} I\right)$. These approximations are only valid if they are sensible initially in the system $(t=0)$ and if $N \gg d_{w} .$ Different approximation is similar for different graphs and is done in (1).\\
\\
In order to test the validity of our simulations, we perform stochastic simulations of the models and compare them with solutions to our system of differential equations. Label first the edges of $(V, H \cup W=E)$ as $e_{i}$ and nodes as $v_{i}$, and we define its incidence matrix $J$ as follows:
$$
J_{i j}= \begin{cases}1 & \text { if } v_{i} \in e_{j} \\ 0 & \text { otherwise }\end{cases}
$$
Let $x(t)=\left(x_{1}(t), \ldots, x_{N}(t)\right)$ be a vector such that $x_{i}(t)=1$ if $v_{i}$ is infected at time $t$ and 0 otherwise. Define the vector $y(t)$ similarly except that $y_{i}(t)=1$ if node $v_{i}$ is susceptible at time $t$. Then $(x J)_{j}\left((y J)_{j}\right)$ gives the number of infected (susceptible) nodes in hyperedge $e_{j}$. The probability of infection of node $v_{i}$ in a time $\delta t$ is then $1-\exp \left(-\tau \sum_{e_{i} \in E} J_{i j} f\left((x J)_{j}\right) \delta t\right.$. We then simulate transitions at each step by generating a random $N$ dimensional vector $r$ such that each component $r_{i} \in(0,1)$. In a time $\delta t$ an infection occurs at $v_{i}$ if $r_{i}<1-\exp \left(-\tau \sum_{e_{i} \in E} J_{i j} f\left((x J)_{j}\right) \delta t\right)$ and $y_{i}=1$; a recovery occurs if $r_{i}<1-\exp (-\gamma \delta t)$, and $x_{i}=1$; a deactivation occurs if $r_{i}<2-\exp (-\beta \delta t)+\exp (-\gamma \delta t)$, $r_{i}>1-\exp (-\gamma \delta t)$ and $x_{i}=1$. The number of deactivated nodes is then $N-\sum_{i} x_{i}(t)+y_{i}(t)$. This generates a statistically accurate distribution []. Multiple simulations are carried out with identical initial conditions ( with constant $\delta t$ and length) and averaged to find $[I],[S]$ and $[D]$. We can make this simulation arbitrarily accurate by taking $\delta t \rightarrow 0$.\\
\\
\textbf{Results}\\
\\
One important thing to note is that in (1) the model was not too accurate at early times. This could be because the initial number of infected nodes was not uniformly distributed. Indeed when we compare 1 and 2 , we see that the predictions are significantly more accurate when the initial conditions are uniformly distributed. This is because in 23 we made the approximation that each hyperedge of a certain type contained the same number of infected nodes. Comparing 3 with 1 and 2 we note that accuracy is significantly reduced when $d_{h}, d_{w} \ll N$. This is to be expected, as in this regime, it is unlikely that each hyper edge of each type has the same number of infected nodes. However when $d_{h}, d_{w}$ are larger, this approximation is more likely to be accurate [4]. Decreasing the value of $\tau$ shifts the location of the peak to a lower time and flattens it as expected 7 . Taking $\gamma=0$ we recover the $S I R$ model whose profile in 6 looks very similar to standard SIR models. Similarly, taking $\beta=0$ we recover the SIS model 5. However for very small values of $\beta$ do not seem to approach the SIS model. If we look at the steady state solutions of $15,16,17$, we can see that $[I]=0,[S]=0,[D]=N$ always no matter what positive values of $\beta, \gamma$ are chosen. This indicates a discontinuity in the phase space of this system.\\
\\
\includegraphics[scale=0.5]{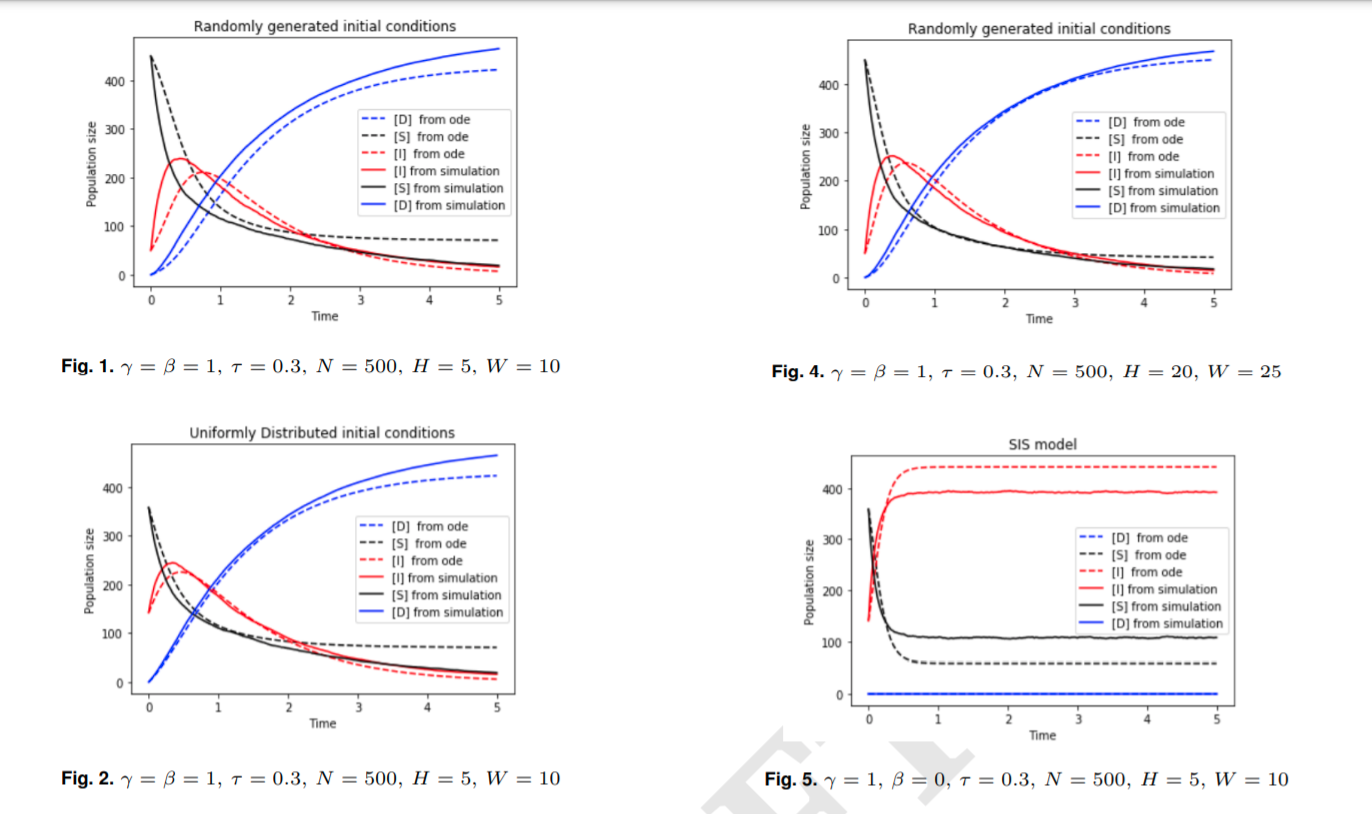}
\begin{center}
\includegraphics[scale=0.5]{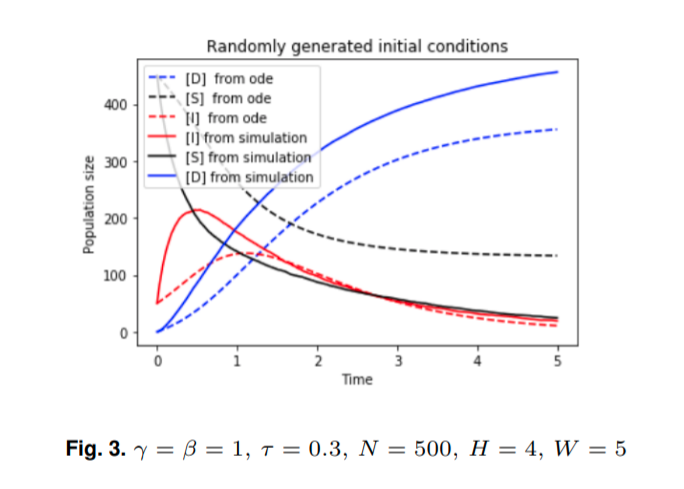}[h!]
\end{center}
\begin{center}
\includegraphics[scale=0.8]{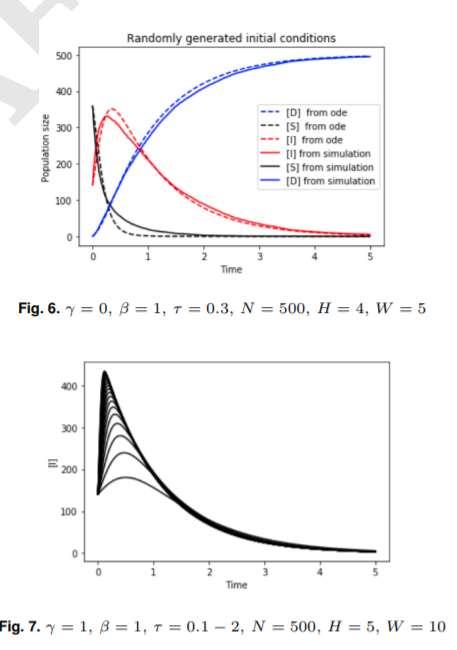}[h!]
\end{center}
\textbf{Discussion}\\
\\
In this report, we generalised the SIS model on hypergraphs to a broader class of spreading processes. The exact equations for the time evolution of expected values were derived in a similar way to (1). A method to simulate the corresponding stochastic process as described and numerical simulations were compared to these. We found strong agreement was found as long as the approximations and initial conditions were compatible and the size of each hyperedge wasn't small when compared to the number of nodes.\\
\\
Our model predicts correct profiles but are only accurate in very specific regimes due to our approximation being valid only in those regimes. All the dependencies of our model were only hypothesized based on numerical simulations. A proper phase space analysis is required to further investigate the possible regimes of SISR model. Also, we have only looked at approximations for one type of hypergraphs being used and these can perhaps be generalised for hypergraphs with an arbitrary distribution of nodes (i.e. by characterising the nodes hypergraph in terms of the probability that they belong to some hyperedge). In order to apply this model to realistic situations, the function $f$ would need to be fitted with known data. Further depending on the needs, the parameters $\gamma, \beta$ should potentially depend on other factors such that how many times a given node has been infected. The advantage of the SISR model is that it takes into account that not all elements transition into a 3rd state instantly. The model also accounts for community structure as well as non-linear dependence of infection pressure on the number of neighbours in each edge.\\
\\
\textbf{References.}\\
\\
1. GYK Ágnes Bodó, Besson, PL Simon, Sis epidemic propagation on hypergraphs. Bull. Math. Biol. 78, 713-735 (2016).\\
\\
2. PV Mieghem, Exact markovian sir and sis epidemics on networks and an upper bound for the epidemic threshold (2014).\\
\\
3. GD Prato, Kolmogorov Equations for Stochastic PDEs. (Birkhãuser, Basel), (2004).
\end{document}